\documentclass[final]{aipproc}

\layoutstyle{8x11single}

\usepackage[latin1]{inputenc}
\usepackage[english]{babel}
\usepackage{bm}
\usepackage{mathrsfs}
\usepackage{color}
\usepackage{xcolor}
\definecolor{magenta}{rgb}{1.2,0.15,0.6}
\definecolor{gold}{rgb}{1,0.8,0.3}
\definecolor{lightblue}{rgb}{0.5,0.8,1.6}
\definecolor{lightblue2}{rgb}{0.5,0.5,1.0}
\definecolor{dark}{rgb}{0.10,0.2,0.3}
\definecolor{light}{rgb}{1.7,1.5,0.6}
\definecolor{light2}{rgb}{1,0.9,0.3}
\definecolor{purpure}{rgb}{0.5,0.15,0.3}
\definecolor{bluelight}{rgb}{0.95,1.4,1.4}
\definecolor{fucsia}{rgb}{0.8,0.1,0.5}
\definecolor{violet}{rgb}{0.5,0.1,0.95}
\definecolor{bluegreen}{rgb}{0,0.6,0.6}
\usepackage{amsmath}
\usepackage{amssymb}

\long\def\comment#1{ }
\newcommand{\bx}{{\bm x}}
\newcommand{\by}{{\bm y}}

\newcommand{\bbx}{{\bar{\bm x}}}
\newcommand{\bby}{{\bar{\bm y}}}

\newcommand{\bz}{{\bm z}}
\newcommand{\bq}{{\bm q}}
\newcommand{\bk}{{\bm k}}

\newcommand{\bu}{{\bm u}}
\newcommand{\bv}{{\bm v}}
\newcommand{\feta}{{\bm \eta}}
\newcommand{\fetab}{{\bar{\bm \eta}}}

\newcommand{\rmd}{{\rm d}}
\newcommand{\rme}{{\rm e}}
\newcommand{\rmi}{{\rm i}}

\newcommand{\rmTr}{{\rm Tr}}

\begin{document}

\title{Particle Production with Rapidity Correlations in Proton-Nucleus Collisions}


\author{E. Iancu}{
  address={Institut de Physique Théorique, CEA Saclay, F-91191 Gif-sur-Yvette, France.}
}

\author{J.D. Madrigal Martínez\footnote{Speaker.}~}{
  address={Institut de Physique Théorique, CEA Saclay, F-91191 Gif-sur-Yvette, France.}
}

\author{D.N. Triantafyllopoulos}{
  address={European Centre for Theoretical Studies in Nuclear Physics and Related Areas (ECT*),
\\\& Fondazione Bruno Kessler, Strada delle Tabarelle 286, I-38123 Villazzano (TN), Italy.}
}

\begin{abstract}
We illustrate how a generalization of the JIMWLK formalism which treats independently the direct and complex conjugate amplitudes can be used to describe semi-inclusive multiparticle production with rapidity correlations in $pA$ collisions. The evolution equations that are obtained with this formalism are susceptible of numerical implementation as a Langevin process.
\end{abstract}

\maketitle

\section{Introduction}

Valuable information about hadronic wavefunctions at high parton densities is provided by the study of multi-particle rapidity correlations in high-energy $pA$ collisions (see, e.g. \cite{Kovnerreview} and references therein). In fact, as a consequence of causality, such correlations must have been produced very shortly after the collision  \cite{causality}. Of course, final state interactions and collective phenomena can also act as a source of correlations between the produced particles and as a matter of fact, these effects are expected to play the key role in describing the ridge observed  in $AA$ collisions at RHIC and LHC. However, the discovery by CMS of ridge-like correlations in $pA$ and even in high multiplicity $pp$ collisions at 7 TeV \cite{CMS}, for which one expects no flow since the size of the interaction region and the number of produced particles are considerably smaller, challenges an explanation based on final state effects. In order to disentangle the correlations produced by initial and final state interactions, it is necessary to be able to compute the cross section for semi-inclusive multiparticle production production in dilute-dense collisions from first principles.\\

An inherent complication of these calculations, as opposed to the computation of total inclusive cross sections, is the need to track independently the direct and complex conjugate amplitudes. Yet another major difficulty  is the absence of $k_T$ factorization for multiparticle production with rapidity separations in the presence of multiple scattering. This problem does not appear in the vast majority of production processes in DIS or $pA$ collisions studied in the literature for which multiple scattering is included in the eikonal approximation, like inclusive gluon production in quark (proton)-nucleon collisions and single inclusive quark valence \cite{12}, prompt photon and Drell-Yan dilepton \cite{3}, inclusive gluon-gluon and gluon-valence quark \cite{4,marquet} or inclusive $q\bar{q}$ production \cite{5}, since such problems either refer to single-inclusive gluon production with an unmeasured projectile, in which case $k_T$ factorization is recovered as a consequence of the cancellation of final state interactions \cite{chen}, or to 2-particle production with both particles propagating at forward rapidities. \\

To deal with these difficulties a suitable generalization of the JIMWLK evolution Hamiltonian was proposed in \cite{formalism,formalism2}, which acts onto a generating functional expressing the doubling of degrees of freedom which are treated independently for the amplitude and its complex conjugate. In the following we review how particle production fits within this picture and we illustrate in the simplest case how this formalism can be applied to compute particle production at a rapidity separation in dilute-dense scattering \cite{preparation}.

\section{The Generating Functional and the Production Hamiltonian}

Consider the gluon-initiated production of a gluon with definite momentum $k$ in the presence of a background field. In the eikonal approximation, the differential cross section is given by\footnote{For simplicity, we are taking the large-$N_c$ limit in Eqs.~\eqref{prod1} and \eqref{prod2}. The evolution and production Hamiltonians, however, are of course not restricted to such a limit.} \cite{marquet}
\begin{equation}\label{prod1}
 \begin{aligned}
  \frac{\rmd\sigma^{g(p)A\to g(q)g(k)X}}{\rmd^2\bk\rmd^2\bq \rmd y_k\rmd y_q}&=q^+\delta(p^+-q^+)\frac{\alpha_s C_A}{\pi^2}\int_{\bx\bbx\by\bby}{\cal K}^i_{\bx\by}{\cal K}^i_{\bbx\bby}\rme^{-\rmi\bq\cdot (\bx-\bbx)-\rmi\bk\cdot(\by-\bby)}\\&\times\left[\langle S_{\by\bby}\rangle_Y \langle S_{\bby\by}\rangle_Y-\langle S_{\by\bbx}\rangle_Y \langle S_{\bbx\bby}\rangle_Y \langle S_{\bby\by}\rangle_Y-\langle S_{\bby\bx}\rangle_Y \langle S_{\bx\by}\rangle_Y \langle S_{\by\bby}\rangle_Y+\langle Q_{\bx\by\bby\bbx}\rangle_Y \langle S_{\bx\bbx}\rangle_Y\langle S_{\by\bby}\rangle_Y\right], 
 \end{aligned}
\end{equation}
where $S_{\bx\by}\equiv\frac{1}{N_c}\rmTr(V^\dagger_\bx V_\by)$ and $Q_{\by\bx\bbx\bby}\equiv\frac{1}{N_c}\rmTr ( V^\dagger_\by V_\bx V^\dagger_\bbx V_\bby)$, with $V_{\bx_i}^\dagger\equiv {\cal P}\exp \left[\rmi g\int\rmd x^+{\cal A}^-_a(x^+,0,\bx_i)t^a\right]$ a Wilson line describing a right-moving projectile in the fundamental representation (we will denote correspondingly by the letter $U$ the Wilson lines in the adjoint representation). We have also introduced the notation $ {\cal K}_{\bx\by\bz}={\cal K}^i_{\bx\bz}{\cal K}^i_{\by\bz};\,\,{\cal K}^i_{\bx\bz}=\frac{(\bx-\bz)^i}{(\bx-\bz)^2}$ for the Weizsäcker-Williams kernel.\\

\begin{figure}[ht]
\includegraphics[width=0.6\textwidth]{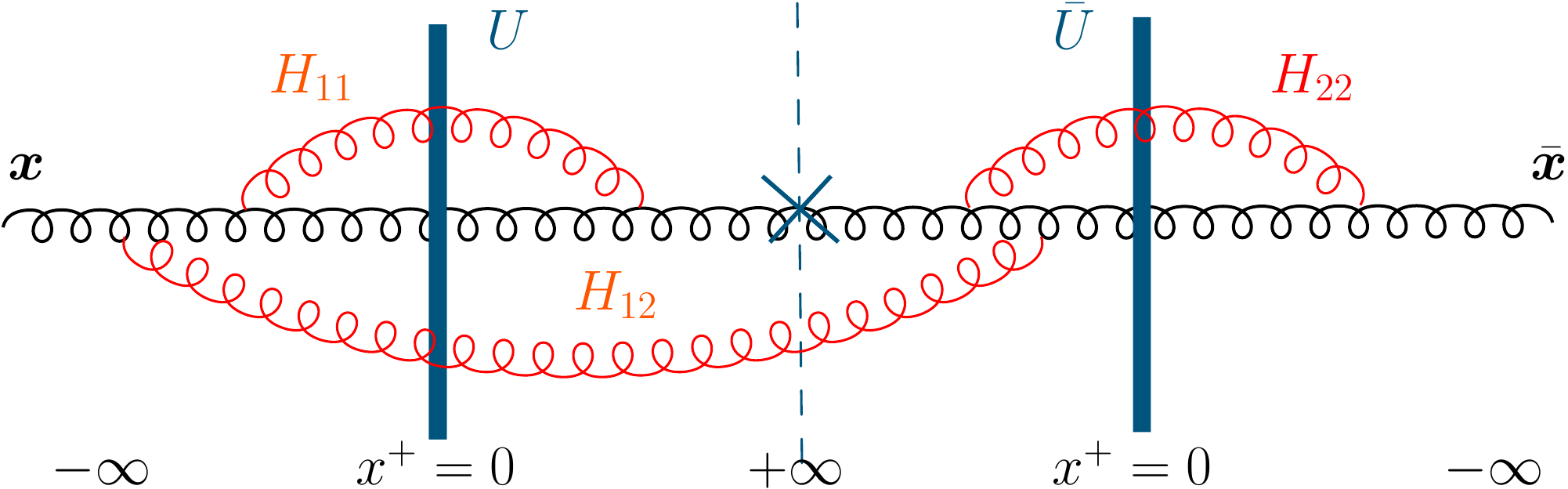}
\caption{Graphs associated to different pieces of the evolution Hamiltonian.}
\label{fig1}
\end{figure}

The differential cross-section \eqref{prod1} can be also expressed as obtained by the action of the production Hamiltonian\footnote{The right derivatives in \eqref{hprod} are defined through their action on Wilson lines, $R^a_\bu V^\dagger_\bx={\rm i}g\delta_{\bu\bx}V_\bx^\dagger t^a $.}
\begin{equation}\label{hprod}
 H_{\rm prod}(\bk)=\frac{1}{4\pi^3}\int_{\by\bby}\rme^{-\rmi\bk\cdot(\by-\bby)}\int_{\bu\bv}{\cal K}^i_{\by\bu}{\cal K}^i_{\bby\bv}(U^\dagger_\bu-U^\dagger_\by)^{ca}(\bar{U}^\dagger_\bv-\bar{U}^\dagger_{\bby})^{cb}R^a_\bu\bar{R}^b_\bv
\end{equation}
onto the gluon generating functional $\hat{S}^A_{\bx\bbx}=\frac{1}{N_c^2-1}\rmTr[U^\dagger_\bx\bar{U}_{\bbx}]$ \cite{formalism,formalism2}:
\begin{equation}\label{prod2}
 \frac{\rmd\sigma^{g(p)A\to g(q)g(k)X}}{\rmd^2\bk\rmd^2\bq\rmd y_k\rmd y_q}=q^+\delta(p^+-q^+-k^+)\frac{1}{(2\pi)^4}\int_{\bx\bbx}\rme^{-\rmi\bq\cdot(\bx-\bbx)}\langle H_{\rm prod}(\bk)\hat{S}^A_{\bx\bbx}|_{U=\bar{U}}\rangle_Y.
\end{equation}
The production Hamiltonian \eqref{hprod} treats differently Wilson lines in the direct ($U$) and complex conjugate amplitudes ($\bar{U}$), and is used to produce on-shell gluons with a definite momentum. One can also evolve the system in rapidity by emitting gluons that are not measured in the final state, by means of the evolution Hamiltonian (Fig.~\ref{fig1})
\begin{equation}
\begin{aligned}
 H_{\rm evol}&=H_{\rm JIMWLK}[U,R;U,R]+H_{\rm JIMWLK}[U,R;\bar{U},\bar{R}]+H_{\rm JIMWLK}[\bar{U},\bar{R};U,R]+H_{\rm JIMWLK}[\bar{U},\bar{R};\bar{U},\bar{R}]\\&\equiv H_{11}+H_{12}+H_{21}+H_{22};\\& H_{\rm JIMWLK}[U,R;\bar{U},\bar{R}]=\frac{1}{(2\pi)^3}\int_{\bu\bv\bz}{\cal K}_{\bu\bv\bz}(U^\dagger_\bu-U^\dagger_\bz)^{ab}(\bar{U}^\dagger_\bv-\bar{U}^\dagger_\bz)^{ac}R^b_\bu\bar{R}^c_\bv.
 \end{aligned}
\end{equation}
A general semi-inclusive multiproduction cross-section in the eikonal approximation can be written in the form
\begin{equation}\label{generic}
\begin{aligned}
 &\frac{\rmd\sigma^{{\cal P}(p_1,\cdots,p_n)A\to {\cal P}(q_1,\cdots,q_n)g(k_1,\Delta Y_1)\cdots g(k_m,\Delta Y_m)X}}{\rmd^2\bq_1\rmd y_{q_1}\cdots\rmd^2\bq_n\rmd y_{q_n}\rmd^2\bk_1\rmd y_{k_1}\cdots\rmd^2\bk_m\rmd^2 y_{k_m}}=\frac{\prod_{i=1}^n p_i^+\delta(q_i^+-p_i^+)}{(2\pi)^{4+2n}}\int_{\bx_1\bbx_1\cdots\bx_n\bbx_n}\rme^{-\rmi[\bq_1\cdot(\bx_1-\bbx_1)+\cdots +\bq_n\cdot(\bx_n-\bbx_n)]}\\&\qquad\qquad\times \bigg\langle H_{\rm prod}(\bk_m)\rme^{H_{\rm evol}\Delta Y_m} H_{\rm prod}(\bk_{m-1})\rme^{H_{\rm evol}\Delta Y_{m-1}}\cdots\times H_{\rm prod}(\bk_1)e^{H_{\rm evol}\Delta Y_1}{\cal Z}_{\bx_i,\bbx_i}|_{U=\bar{U}}\bigg\rangle_Y,
 \end{aligned}
\end{equation}
where ${\cal Z}_{\bx_i,\bbx_i}$ is the generating functional (density matrix) associated to the projectile, which consists of Wilson lines with coordinates $\bx_i$ in the direct amplitude and $\bbx_i$ in the complex conjugate ($i=1,\cdots,n$). Notice that in the spirit of collinear factorization we are taking the transverse momenta of the components of the incoming projectile to be zero. It is important to remark that in view of Eq.~\ref{generic}, no factorization in rapidity is to be expected except for very special cases, since the projectile evolution is entangled with multiple scattering from the target in a rather nontrivial fashion.\\

The concrete incarnations of the general formula \eqref{generic} rapidly become very involved. In order to be able to act sequentially with $H_{\rm evol}$ and $H_{\rm prod}$,  it is needed to conserve the functional dependence of the Wilson lines at each step of the evolution, which introduces a tremendous complexity. Fortunately, in Ref.~\cite{formalism2}, a numerics-friendly procedure was devised along the lines of \cite{path} to cast the evolution as a bilocal Langevin process.\\

\begin{figure}[ht]
\includegraphics[width=0.7\textwidth]{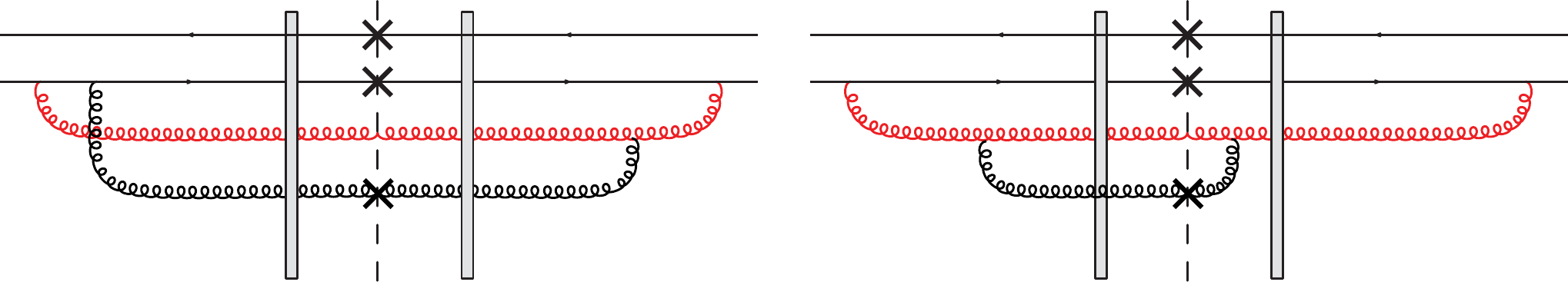}
\caption{A sample of typical shockwave diagrams for gluon production in a dipole-nucleus collision with one step in rapidity evolution (red gluon). The crosses indicate those partons whosee momentum is measured.}
\label{fig2}
\end{figure}
\vspace{-0.2cm}
We note that, when the components of the projectile do not get their transverse momenta measured, only the right derivative terms turn out to be relevant for its rapidity evolution due to causality (see, for instance, the discussion in \cite{formalism2}). In such a case, the single-inclusive gluon production by a generic projectile described by the generating functional ${\cal Z}_{\Delta Y}=\rme^{H_{\rm evol}\Delta Y}{\cal Z}_{Y=0}$ (where $\Delta Y$ is the rapidity difference between the valence components of the projectile and the produced gluon) can be computed as
\begin{equation}\label{factor}
\begin{aligned}
 \frac{\rmd\sigma^{{\cal P}A\to g(k)X}}{\rmd^2\bk\rmd y_k}&=\frac{1}{(2\pi)^4}\frac{1}{4\pi^3}\int_{\by\bby}\rme^{-\rmi\bk\cdot(\by-\bby)}\int_{\bu\bv}{\cal K}^i_{\by\bu}{\cal K}^i_{\bby\bv}\big\langle(U^\dagger_\bu-U^\dagger_\by)^{ca}(U^\dagger_\bv-U^\dagger_\bby)^{cb}\big\rangle_Y [R^a_\bu \bar{R}^b_\bv{\cal Z}_{\Delta Y}]_{V=\bar{V}}.
 \end{aligned}
\end{equation}
Notice that we must have $[R^a_\bu \bar{R}^b_\bv{\cal Z}_{\Delta Y}]_{V=\bar{V}}=\frac{\delta^{ab}}{N_c^2-1}R^e_\bu \bar{R}^e_\bv{\cal Z}_{\Delta Y}|_
{V=\bar{V}}\equiv -\delta^{ab}n_{\Delta Y}(\bu,\bv)$, where the charge correlator $n_{\Delta Y}(\bu,\bv)$ appears outside the target average since it is not dependent on the Wilson lines after setting $V=\bar{V}$ (this is also the reason why it is color diagonal). Then we have a similar situation to the $k_T$-factorization of BFKL physics \cite{book} and it is natural to identify $n_Y$ as the unintegrated gluon distribution (see also next section).

\section{Non-Forward Gluon Production in Dipole-Nucleus Scattering}

The formalism introduced in the last section permits not only to easily recover the cross-sections for particle production of Refs.~\cite{12,3,4,5}, but also to address the interesting and more difficult case where the momenta of the projectile components is measured and rapidity evolution is allowed between the projectile and the identified gluon. The simplest process of this kind is gluon production out of a dipole ---in which the momenta of the quark and the antiquark are measured---, which has evolved one step in rapidity (Fig. \ref{fig2}).  According to \eqref{generic}, we will have
\begin{equation}\label{fis}
 \begin{aligned}
  \frac{\rmd\sigma^{q(p_1)\bar{q}(p_2)A\to q(q_1)\bar{q}(q_2)g(k,\delta Y)X}}{\rmd^2\bq_1\rmd y_{q_1}\rmd^2\bq_2\rmd y_{q_2}\rmd^2\bk\rmd y_k}&=\frac{\prod_{i=1}^2p_i^+\delta(p_i^+-q_i^+)}{(2\pi)^8}\int_{\bx\bbx\by\bby}\rme^{-\rmi[\bq_1\cdot (\bx-\bbx)+\bq_2\cdot(\by-\bby)]}\big\langle H_{\rm prod}(\bk)Q^{(1)}_{\bx\by\bby\bbx}|_{V=\bar{V}}\big\rangle_Y,
 \end{aligned}
\end{equation}
where $Q^{(1)}_{\bx\by\bby\bbx}$ is the dipole generating functional after evolving one step $\delta Y$ in rapidity through the emission of an unmeasured real or virtual gluon, given by \cite{formalism2}

\begin{equation}\label{qevol}
 \begin{aligned}
 Q^{(1)}_{\bx\by\bby\bbx}=&Q_{\bx\by\bby\bbx}+\frac{\bar{\alpha}\,\delta Y}{4\pi}\int_\bz \big\{({\cal M}_{\bx\by\bz}+{\cal M}_{\bx\bbx\bz}-{\cal M}_{\bbx\by\bz})S_{\bx\bz}Q_{\bz\by\bby\bbx}+({\cal M}_{\bx\by\bz}+{\cal M}_{\by\bby\bz}-{\cal M}_{\bx\bby\bz}) S_{\bz\by}Q_{\bx\bz\bby\bbx}\\+&({\cal M}_{\bbx\bby\bz}+{\cal M}_{\bx\bbx\bz}-{\cal M}_{\bx\bby\bz}) \bar{S}_{\bz\bbx}Q_{\bx\by\bby\bz}+({\cal M}_{\bbx\bby\bz}+{\cal M}_{\by\bby\bz}-{\cal M}_{\bbx\by\bz})\bar{S}_{\bby\bz}Q_{\bx\by\bz\bbx}-({\cal M}_{\bx\by\bz}+{\cal M}_{\bbx\bby\bz}+{\cal M}_{\bx\bbx\bz}+{\cal M}_{\by\bby\bz}) Q_{\bx\by\bby\bbx}\\-&({\cal M}_{\bx\by\bz}+{\cal M}_{\bbx\bby\bz}-{\cal M}_{\bbx\by\bz}-{\cal M}_{\bx\bby\bz}) S_{\bx\by}\bar{S}_{\bby\bbx}-({\cal M}_{\bx\bbx\bz}+{\cal M}_{\by\bby\bz}-{\cal M}_{\bbx\by\bz}-{\cal M}_{\bx\bby\bz}) Q_{\bx\bz\bz\bbx}Q_{\bz\by\bby\bz}\big\},
 \end{aligned}
\end{equation}
where $\bar{S}_{\bbx\bby}=\frac{1}{N_c}\rmTr[\bar{V}_{\bbx}\bar{V}^\dagger_{\bby}]$ and $ {\cal M}_{\bx\by\bz}={\cal K}_{\bx\bx\bz}+{\cal K}_{\by\by\bz}-2{\cal K}_{\bx\by\bz}$ is the dipole kernel. The final expression for the cross section \eqref{fis} is rather lengthy even in the large-$N_c$ limit \cite{preparation}, hence we will not reproduce it here. Interestingly, it gets dramatically simplified when we do not measure the projectile partons, i.e. we integrate over $q_1$ and $q_2$. Then it is easy to see that we regain the factorized expression \eqref{factor}
\begin{equation}
 \begin{aligned}
  \frac{\rmd\sigma^{q\bar{q}A\to g(k,\delta Y)X}}{\rmd^2\bk\rmd y_k}&=-\frac{1}{(2\pi)^4}\int_{\feta\fetab}\frac{\rme^{-\rmi\bk\cdot(\feta-\fetab)}}{4\pi^3}\int_{\bu\bv}{\cal K}_{\feta\bu}^i{\cal K}^i_{\fetab\bv}(S_{\bu\bv}S_{\bv\bu}-S_{\feta\bv}S_{\bv\feta}-S_{\fetab\bu}S_{\bu\fetab}+S_{\feta\fetab}S_{\fetab\feta})n_{\delta Y}(\bu,\bv|\bx,\by),\\
n_{\delta Y}(\bu,\bv|\bx,\by)&\equiv -\frac{1}{N_c^2-1} R^a_{\bu}\bar{R}^a_{\bv}Z_{\delta Y}(\bx,\by)|_{V=\bar{V}};\quad Z_Y(\bx,\by)\equiv \rme^{H_{\rm evol}Y}Q_{\bx\by\by\bx}.
\end{aligned}
\end{equation}
The evolution of $Z_Y(\bx,\by)$ follows a simplified version of \eqref{qevol}, which is in fact familiar from the dipole picture \cite{dipole}
\begin{equation}
 \frac{\partial Z_Y(\bx,\by)}{\partial Y}=-\frac{\bar{\alpha}}{2\pi}\int_\bz {\cal M}_{\bx\by\bz}[Z_Y(\bx,\by)-Z_Y(\bx,\bz)Z_Y(\bz,\by)].
 \end{equation}
 Since $Z_Y[V,V]=1$ and $R^aZ_Y[V,\bar{V}]|_{V=\bar{V}}=0$, we are driven to 
\begin{equation}\label{eq}
 \frac{\partial n_Y(\bu,\bv|\bx,\by)}{\partial Y}=-\frac{\bar{\alpha}}{2\pi}\int_\bz{\cal M}_{\bx\by\bz}[n_Y(\bu,\bv|\bx,\bz)+n_Y(\bu,\bv|\bz,\by)-n_Y(\bu,\bv|\bx,\by)],
\end{equation}
which we recognize as the BFKL equation. This supports the identification of $n_Y$ as the unintegrated gluon distribution.
\vspace{-0.2cm}
\begin{theacknowledgments}
 J.D.M. would like to thank the organizers of DIFFRACTION 2014 for the pleasant atmosphere in which the workshop took place. Our research is supported by the Agence Nationale de la Recherche under the project \# 11-BS04-015-01 and by the European Research Council under the Advanced Investigator Grant ERC-AD-267258.
\end{theacknowledgments}


\begin{thebibliography}{17}
\expandafter\ifx\csname natexlab\endcsname\relax\def\natexlab#1{#1}\fi
\providecommand{\enquote}[1]{``#1''}
\expandafter\ifx\csname url\endcsname\relax
  \def\url#1{\texttt{#1}}\fi
\expandafter\ifx\csname urlprefix\endcsname\relax\def\urlprefix{URL }\fi
\providecommand{\eprint}[2][]{\url{#2}}




\bibitem[Kovner (2013)]{Kovnerreview}
A.~Kovner, and M.~Lublinsky, \emph {Int.J.Mod.Phys.} {\bf E22}, 1330001 (2013).
 
 \bibitem[Dumitru (2008)]{causality}
A.~Dumitru, F.~Gelis, L.~McLerran, and R.~Venugopalan,  \emph{Nucl.Phys.} {\bf A810}, 91 (2008).

\bibitem[CMS (2010)]{CMS}
V.~Khachatryan, et al. [{\bf CMS Collaboration}], \emph{JHEP} {\bf 1009} (2010) 091.

\bibitem[Kovchegov (1998)]{12}
Yu. V. Kovchegov, and A. H. Mueller, \emph{Nucl.Phys.} {\bf B529}, 451 (1998); B. Z. Kopeliovich, A. V. Tarasov, and A. Schäfer, \emph{Phys.Rev.} {\bf C59}, 1609 (1999); A. Dumitru, and L. D. McLerran, \emph{Nucl.Phys.} {\bf A700}, 492 (2002); Yu. V. Kovchegov, and K. Tuchin, \emph{Phys.Rev.} {\bf D65}, 074026 (2002).

\bibitem[Gelis (2002)]{3}
F. Gelis, and J. Jalilian-Marian, \emph{Phys.Rev.} {\bf D66}, 014021 (2002); \emph{Phys.Rev.} {\bf D66}, 094014 (2002); B. Z. Kopeliovich, V. Raufeisen, A. V. Tarasov, and M. B. Johnson, \emph{Phys.Rev.} {\bf C67}, 014903 (2003); R. Baier, A.H. Mueller, and D. Schiff, \emph{Nucl.Phys.} {\bf A741}, 358 (2004).

\bibitem[Marian (2004)]{4}
J. Jalilian-Marian, and Yu. V. Kovchegov, \emph{Phys.Rev.} {\bf D70}, 114017 (2004); R. Baier, A. Kovner, M. Nardi, and U. A. Wiedemann, \emph{Phys.Rev.} {\bf D72}, 094013 (2005); A. Kovner, and M. Lublinsky, \emph{JHEP} {\bf 0611}, 083 (2006); E. Iancu, and J. Laidet, \emph{Nucl.Phys.} {\bf A916}, 48 (2013).

\bibitem[Marquet (2007)]{marquet}
C. Marquet, \emph{Nucl.Phys.} {\bf A796}, 41 (2007); J. L. Albacete, and C. Marquet, \emph{Phys.Rev.Lett.} {\bf 105}, 162301 (2010).

\bibitem[Blaizot (2004)]{5}
J.-P. Blaizot, F. Gelis, and R. Venugopalan, \emph{Nucl.Phys.} {\bf A743}, 57 (2004); Yu. V. Kovchegov, and K. Tuchin, \emph{Phys.Rev.} {\bf D74}, 054014 (2006).

\bibitem[Chen (1995)]{chen}
Z. Chen, and A. H. Mueller, \emph{Nucl.Phys.} {\bf B451}, 579 (1995).

\bibitem[Hentschinski (2006)]{formalism}
M. Hentschinski, H. Weigert, and A. Schäfer, \emph{Phys.Rev.} {\bf D73}, 051501 (2006); A. Kovner, M. Lublinsky, and H. Weigert, \emph{Phys.Rev.} {\bf D74}, 114023 (2006).

\bibitem[Iancu (2013)]{formalism2}
E. Iancu, and D. N. Triantafyllopoulos, \emph{ JHEP} {\bf 1311}, 067 (2013).

\bibitem[Iancu (2014)]{preparation}
E. Iancu, J. D. Madrigal, and D. N. Triantafyllopoulos, \emph{ in preparation}.

\bibitem[Blaizot (2003)]{path}
J.-P. Blaizot, E. Iancu, and H. Weigert, \emph{Nucl.Phys.} {\bf A713}, 441 (2003).

\bibitem[Kovchegov (2012)]{book}
Yu. V. Kovchegov, and E. Levin, \emph{Quantum Chromodynamics at High Energy}, Cambridge University Press, Cambridge, 2012.

\bibitem[Mueller (1994)]{dipole}
A. H. Mueller, \emph{Nucl.Phys.} {\bf B415}, 373 (1994).
\end{thebibliography}
\end{document}